 \documentclass[final,5p,times,twocolumn]{elsarticle}

\usepackage{amssymb}
\journal{Physica B}

\begin{document}

\begin{frontmatter}



\title{Competing Hydrostatic Compression Mechanisms in Nickel Cyanide}


\author[ox,es]{J.~Adamson}
\author[bi]{T.~C.~Lucas}
\author[ox]{A.~B.~Cairns}
\author[ox]{N.~P.~Funnell}
\author[is,di]{M.~G.~Tucker}
\author[di]{A.~K.~Kleppe}
\author[bi]{J.~A.~Hriljac}
\author[ox]{A.~L.~Goodwin}

\address[ox]{Department of Chemistry, University of Oxford, Inorganic Chemistry Laboratory, South Parks Road, Oxford OX1 3QR, U.K.}
\address[es]{National Institute of Chemical Physics and Biophysics, Akadeemia tee 23, 12618 Tallinn, Estonia}
\address[bi]{School of Chemistry, University of Birmingham, Edgbaston, Birmingham B15 2TT, U.K.}
\address[is]{ISIS Facility, Rutherford Appleton Laboratory, Harwell Oxford, Didcot, Oxfordshire OX11 0QX, U.K.}
\address[di]{Diamond Light Source, Chilton, Oxfordshire, OX11 0DE, U.K.}

\begin{abstract}
We use variable-pressure neutron and X-ray diffraction measurements to determine the uniaxial and bulk compressibilities of nickel(II) cyanide, Ni(CN)$_2$. Whereas other layered molecular framework materials are known to exhibit negative area compressibility, we find that Ni(CN)$_2$ does not. We attribute this difference to the existence of low-energy in-plane tilt modes that provide a pressure-activated mechanism for layer contraction. The experimental bulk modulus we measure is about four times lower than that reported elsewhere on the basis of density functional theory methods [{\it Phys.~Rev.~B} {\bf 83}, 024301 (2011)].
\end{abstract}


\end{frontmatter}


\section{Introduction}

From a lattice dynamical perspective, molecular frameworks have provided an extraordinary number of interesting examples that challenge our understanding of how materials should respond to external stimuli such as temperature and pressure. Negative thermal expansion in the Zn(CN)$_2$ family \cite{Williams_1997,Goodwin_2005}, pressure-induced and thermal amorphisation in metal--organic frameworks (MOFs) such as ZIF-8 \cite{Chapman_2009,Bennett_2010}, and ``breathing'' transitions in the MIL-53 and MIL-101 systems are all popular examples \cite{Serre_2007,Ma_2012}. A related phenomenon to receive particular recent interest is that of negative compressibility, whereby a material actually expands in one or more directions under application of hydrostatic pressure \cite{Baughman_1998,Cairns_2013}. While it is now emerging that negative linear compressibility (NLC) might be somewhat more widespread amongst molecular frameworks and molecular crystals than originally thought \cite{Cairns_2015}, the much rarer property of negative area compressibility (NAC)---where a system reduces its volume by expanding simultaneously in two orthogonal directions---remains known to occur in only a handful of materials \cite{Loa_1999,Hodgson_2014,Cai_2015}. The development of a microscopic understanding of NAC and its lattice dynamical origins is crucial if we are ever to design improved NAC candidates for application in next-generation ferroelectric sensor technology \cite{Baughman_1998}.

\begin{figure}
\begin{center}
\includegraphics{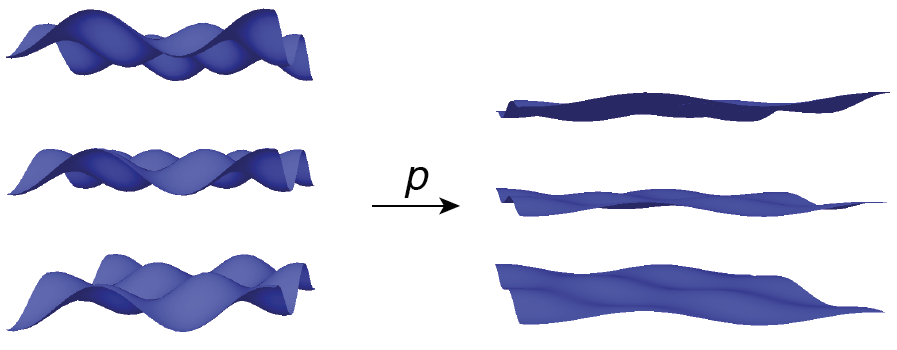}
\end{center}
\caption{The Lifshitz instability in layered materials involves displacement of atoms within layers in a direction parallel to the layer stacking axis. Increased population of Lifshitz modes results in a reduction of the effective cross-sectional area of each layer, which couples to an expansion of the inter-layer separation. The Lifshitz mode is associated with negative area compressibility in cases where volume reduction under hydrostatic pressure results in rapid inter-layer collapse that couples with damping of the Lifshitz modes.\label{fig1}}
\end{figure}

Arguably the clearest lattice dynamical signature of NAC established to date is the existence of so-called Lifshitz modes in layered framework materials. These modes are low-energy vibrations of weakly-coupled layers in which atomic displacements are polarised perpendicular to the layer plane [Fig.~\ref{fig1}]. The relationship with NAC arises from a combination of a large and negative elastic compliance $S_{13}$---linking interlayer compression with layer expansion---and a small value of the elastic constant $C_{33}$ which reflects facile compression along the layer stacking axis. As a consequence, one design strategy for identifying new NAC candidates has been to focus on layered systems with very weak inter-layer interactions, such as is observed in the key NAC system silver(I) tricyanomethanide \cite{Hodgson_2014}.

\begin{figure}
\begin{center}
\includegraphics{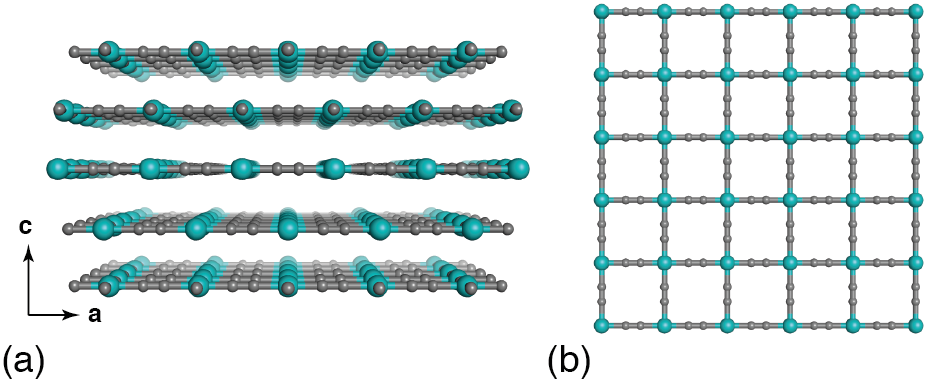}
\end{center}
\caption{Representation of the structure of Ni(CN)$_2$. (a) Square-grid layers are stacked along the $\mathbf c$ crystal axes. Successive layers are shifted by $\frac{1}{2}\langle100\rangle$ lattice vectors. The existence of choice in this shift direction results in a loss of long-range periodicity along the stacking axis. (b) A single layer is assembled from square-planar Ni$^{2+}$ cations (shown here in teal) connected \emph{via} linear Ni--C--N--Ni linkages (C/N atoms shown in grey). The orientation of the cyanide ions is disordered throughout the structure. The view direction in (b) is parallel to the $\mathbf c$ axis.\label{fig2}}
\end{figure}

It was in this context that we sought to study the high-pressure structural behaviour of nickel(II) cyanide, Ni(CN)$_2$. The structure of this material is described by stacked square layers, each assembled from square-planar Ni$^{2+}$ ions connected \emph{via} linear Ni--C--N--Ni linkages [Fig.~\ref{fig2}] \cite{Hibble_2007,Goodwin_2009b}. The interaction between layers is considered to be weak on the basis of a number of observations. First, Ni(CN)$_2$ exhibits strong positive thermal expansion (PTE) behaviour along the layer stacking axis that is coupled to area negative thermal expansion (NTE) within the layer plane \cite{Hibble_2007}; note the magnitude of thermal expansion is usually inversely proportional to the strength of bonding interactions \cite{Ogborn_2012}. Second, there is no long-range periodicity within the stacking arrangement, a structural feature which has also been associated with weak inter-layer interactions \cite{Goodwin_2009b}. And, third, the phonon dispersion curves determined using density functional theory (DFT) calculations reflect the existence of Lifshitz-like excitations \cite{Mittal_2011}. So the established design criteria for NAC in molecular frameworks would suggest that Ni(CN)$_2$ should be a prime candidate in which to observe the unusual phenomenon of NAC.

Here we apply a combination of variable-pressure neutron and X-ray powder diffraction measurements to determine experimentally the compressibility behaviour of Ni(CN)$_2$ over the pressure range $0<p<2.5$\,GPa. We find no evidence for NAC behaviour, a result we rationalise in terms of the existence of low-energy modes with negative Gr{\"u}neisen parameters that compete with the Lifshitz modes in allowing volume collapse via layer densification. Subsequent to carrying out our measurements, a related study has appeared in the preprint literature \cite{Mishra_2014}. The experimental results reported by us and by Ref.~\citenum{Mishra_2014} are broadly similar; however our analysis differs in some important respects. In particular, we interpret our compressibility data in the context of the Gr{\"u}neisen parameters associated with different types of NTE modes supported by the Ni(CN)$_2$ structure. Our own paper is arranged as follows. We begin by describing the experimental methods used in our study. We then present the pressure-dependent lattice parameter variation we observe experimentally and report the corresponding lattice and bulk compressibilities. Our manuscript concludes with a discussion of possible microscopic mechanisms responsible for this behaviour, placing our results in the context of previous studies of Ni(CN)$_2$ and of other NAC materials.

\section{Experimental Methods}

\subsection{Synthesis}

Powder samples of Ni(CN)$_2$ were prepared according to one of two methods. The first method involved heating Ni(CN)$_2\cdot$4H$_2$O (Alfa Aesar, $>$99\%) under vacuum at 200\,$^\circ$C for 24\,h. The dehydration process is coupled to a change in colour; the yellow product was stored in a vacuum desiccator to prevent rehydration. The second method involved reaction of aqueous solutions of nickel(II) nitrate (Sigma Aldrich, $>$99\%) and potassium tetracyanonickelate(II) (Sigma Aldrich, $>$99\%). The precipitate formed was isolated by vacuum filtration, washed (H$_2$O) and dried to afford Ni(CN)$_2\cdot4$H$_2$O. This product was dehydrated and stored as described above.

\subsection{Variable-pressure neutron diffraction}

Neutron diffraction measurements were carried out using the PEARL instrument at the ISIS neutron spallation source. Hydrostatic pressure was applied using a V4 Paris-Edinburgh press. A powdered sample of Ni(CN)$_2$, prepared as described above, was loaded into a TiZr gasket together with a Pb pellet and fluorinert as pressure-transmitting medium. The gasket was positioned within a zirconia-toughened alumina anvil with a Los Alamos single toroidal profile. The sample was subjected to a series of eight force values that corresponded to hydrostatic pressures $0<p<0.78$\,GPa. Pressures were calculated using the third-order Birch-Murnaghan equation of state known for Pb ($V_0=30.354$\,\AA$^3$, $B_0=42.0$\,GPa, $B^\prime=5.71$). Neutron scattering data were collected over the $d$-spacing range $0<d<4.1$\,\AA. Constrained Rietveld refinements were used to extract unit cell parameters via the GSAS refinement package \cite{Larson_2000}. A crystallographic approximant to the ``disorder stack'' model reported in Ref.~\citenum{Goodwin_2009b} was used as the basis of these refinements ($I4_1/amd, a\simeq4.9$\,\AA$, c\simeq12.8$\,\AA), together with a stacking-fault model as implemented in GSAS \cite{Larson_2000}. Our data were not sufficiently discriminating to allow robust refinement of positional parameters.

\subsection{Variable-pressure synchrotron X-ray diffraction}

Synchrotron X-ray diffraction measurements were carried out using the I15 beamline at the Diamond Light Source. A powdered sample of Ni(CN)$_2$, prepared as described above, was loaded under a controlled atmosphere into a modified Merrill-Bassett diamond-anvil cell with diamond culets of 450\,$\mu$m. Silicone oil was used as pressure-transmitting medium. Ruby chips, evenly distributed in the pressure chamber, were used to measure the applied pressure according to the ruby fluorescence method \cite{Piermarini_1975,Mao_1986}. The variation in pressure values determined from different chips was found to be less than 0.1\,GPa. For our measurements, five pressures over the range $0<p<2.5$\,GPa were applied. The observed intensities were integrated as a function of $2\theta$ using the software FIT2D \cite{Fit2D,Hammersley_1996} in order to give one-dimensional diffraction profiles. A powder sample of CeO$_2$ was used to calibrate the beam centre and to determine the sample-to-detector distance. As for the neutron diffraction data, constrained Rietveld refinements were used to extract unit cell parameters via the GSAS refinement package \cite{Larson_2000}. The same structural model described above was used as the basis of these refinements, and again our data were not sufficiently discriminating to allow robust refinement of positional parameters.

\section{Results}

\begin{figure}
\begin{center}
\includegraphics[width=\columnwidth]{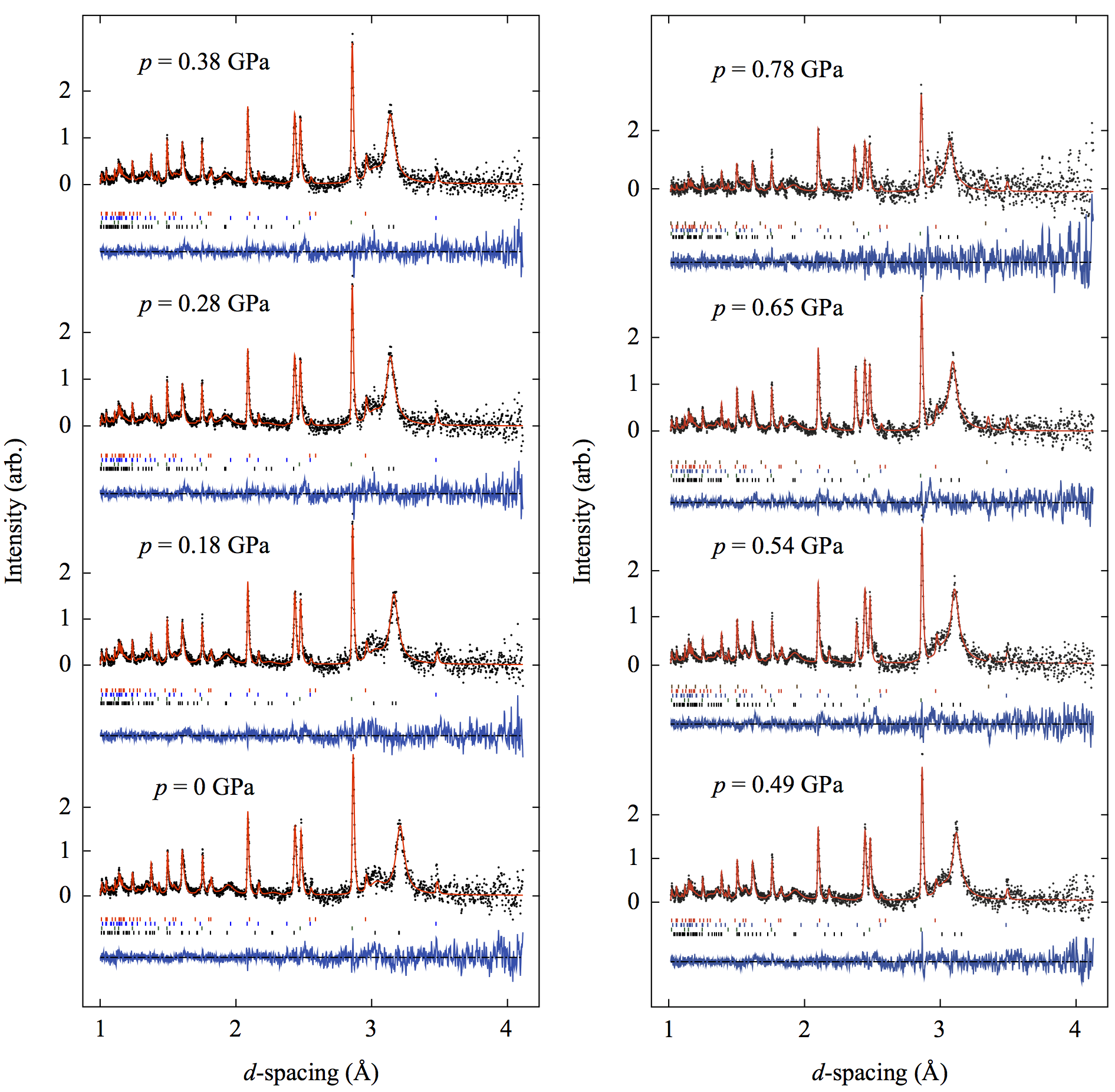}
\end{center}
\caption{Variable-temperature time-of-flight neutron diffraction patterns collected for Ni(CN)$_2$ together with fits obtained from Rietveld refinement. Data are shown as black markers; fit as a red line, and difference (data $-$ fit) in blue. Tick marks are given for the various phases included in the fit: (bottom--top) Ni(CN)$_2$, Pb, Al$_2$O$_3$ and ZrO$_2$ components of the zirconia-toughened alumina (ZTA) anvils, and (optionally) the high-pressure Ni(CN)$_2$ phase.\label{fig3}}
\end{figure}

The time-of-flight neutron diffraction patterns and corresponding fits obtained from Rietveld refinement are shown in Fig.~\ref{fig3}. The coexistence of contrasting sharp and broad reflections noted elsewhere \cite{Goodwin_2009b} as evidence for stacking disorder is clearly visible in our newly-collected data and appears immune to the application of hydrostatic pressure. In contrast to the finding of Ref.~\citenum{Mishra_2014} we do not find any evidence of a structural phase transition below 0.1\,GPa: all data sets collected over the pressure range $0<p<0.5$\,GPa are well fit using a structural model based on the ambient-pressure structure \cite{Goodwin_2009b}. At pressures above 0.5\,GPa we find a set of additional reflections that can be indexed (very tentatively) on the basis of a new Ni(CN)$_2$ cell consistent with a defect Prussian blue architecture ($Im\bar3m, a\simeq4.8$\,\AA). In the absence of any noticeable discontinuity in the compressibility of the ambient phase we consider these additional peaks to arise from a slow first-order transition to a high-pressure phase that may be the $p>0.7$\,GPa phase postulated in Ref.~\citenum{Mishra_2014}. Given that only a handful of relatively-weak peaks contribute to this minor component of the diffraction pattern, we cannot have any real confidence in the structural model we have used for this phase in our Rietveld refinement. Nevertheless, the (meta-)stability of the ambient phase to higher pressures allows us to continue to study its compressibility behaviour over the entire pressure regime $0<p<2.5$\,GPa.

\begin{figure}
\begin{center}
\includegraphics[width=0.6\columnwidth]{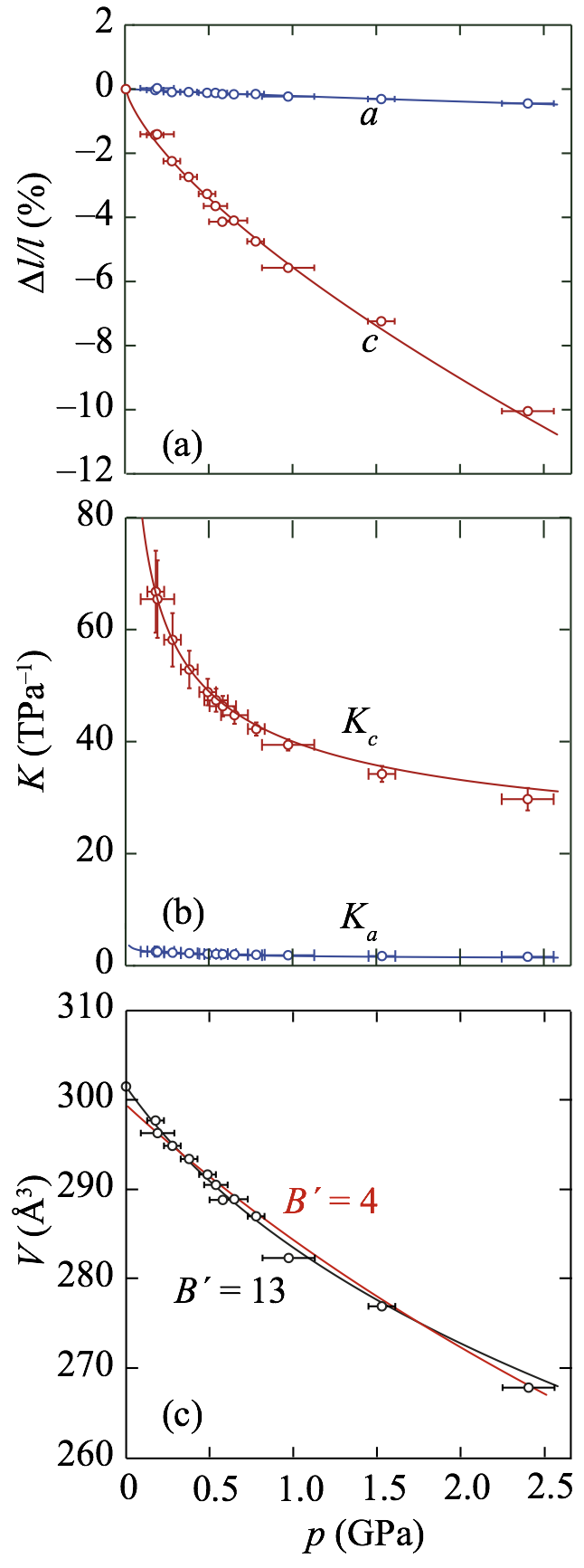}
\end{center}
\caption{Compressibility behaviour of Ni(CN)$_2$. (a) Relative change in lattice parameters under hydrostatic pressure: the mechanical response of the system is dominated by compression along $\mathbf c$. (b) The corresponding linear compressibilities extracted from the empirical fit to the data in (a) according to Eq.~(\ref{eqfit}), (c) The volume equation of state, shown with both second-order ($B^\prime\equiv4$) and third-order Birch-Murnaghan fits as described in the text.\label{fig4}}
\end{figure}

The variation in lattice parameters extracted from our Rietveld fits to these neutron diffraction data---together with those for the X-ray diffraction data (themselves not shown)---is shown in Fig.~\ref{fig4}. What is immediately obvious is that the structural response to hydrostatic pressure is dominated by rapid collapse of the inter-layer separation, which corresponds to the $c$ lattice parameter. By contrast the $a$ parameter, which describes the Ni$\ldots$Ni separation within an individual square sheet, changes remarkably little under hydrostatic pressure. Of particular note is the absence of any increase in this parameter with pressure, which rules out NAC behaviour in Ni(CN)$_2$.

The lattice parameter data were fitted in the program {\sc pasc}al \cite{Cliffe_2012} according to a set of empirical expressions of the form
\begin{equation}
\ell(p)=\ell_0+\alpha(p-p_{\rm c})^\lambda,\label{eqfit}
\end{equation}
as described in Ref.~\citenum{Goodwin_2008c}. The corresponding compressibilities, obtained via the isothermal pressure derivative of Eq.~(\ref{eqfit}), are shown in Fig.~\ref{fig4}(b). Extreme deviations from linear behaviour are observed for the $c$ axis compressibility at low pressures, which suggests strongly anharmonic inter-layer interactions. The linear compressibilities extracted from these fits over the entire pressure range $0<p<2.5$\,GPa are
\begin{eqnarray}
K_a&=&+1.8(2)\,{\rm TPa}^{-1},\\
K_c&=&+45(2)\,{\rm TPa}^{-1}.
\end{eqnarray}
Hence Ni(CN)$_2$ is essentially incompressible within the layer directions and is about as strongly compressible in a direction parallel to the layer stacking axis as the NAC material silver(I) tricyanomethanide ($K_\ell=+66(20)$\,TPa$^{-1}$) \cite{Hodgson_2014}.

A third-order Birch-Murnaghan equation of state \cite{Murnaghan_1944} describes well the pressure-dependence of the unit cell volume [Fig.~\ref{fig4}(c)]. The corresponding zero-pressure bulk modulus is determined as $B_0=11.7(1.8)$\,GPa with a first pressure derivative $B^\prime=12(3)$. Taken together these values reflect a compliant material which rapidly stiffens on application of hydrostatic pressure \cite{Slebodnick_2004}. Our values are consistent with other cyanides and layered molecular frameworks \cite{Hodgson_2014,Collings_2013} but contrast with those reported in Refs.~\citenum{Mittal_2011,Mishra_2014}: $B_0=105(2)$\,GPa ($0<p<0.1$\,GPa; experiment), $15.4(2)$\,GPa ($0.1<p<18$\,GPa; experiment), and $63.4$\,GPa (DFT).

\section{Discussion and Conclusions}

There are two obvious questions posed by these results, and we proceed to answer these in turn. The first is why there might be such a discrepancy between the compressibility values determined here and those reported in Refs.~\citenum{Mittal_2011} and \citenum{Mishra_2014}. Whereas the second---and arguably the more important---concerns why Ni(CN)$_2$ \emph{does not} show NAC when its structure possesses all the design motifs thought to favour the phenomenon. Answering this particular question is key to informing future design strategies for targeting NAC behaviour in other materials.

With respect to the apparent discrepancies in compressibility amongst the various studies of the lattice dynamics of Ni(CN)$_2$, we address first the difference between experimental and computational (DFT) results. The DFT study of Ref.~\citenum{Mittal_2011} was always going to be difficult because the layer stacking sequence in Ni(CN)$_2$ has no true periodicity along the stacking axis \cite{Goodwin_2009b}, and so the material structure is inconsistent with the presumption of crystallinity inherent to DFT methods. The work-around devised in that study was to use a modified layer stacking arrangement in which the layers were separated by twice their normal distance and placed directly above one another rather than offset by $\frac{1}{2}\langle100\rangle$ as known to occur in the real material. This meant that cell relaxation was not possible and the all-important compressibility in the $c$ direction determined computationally is unlikely to have any physical meaning. Even if a different stacking arrangement had been implemented, it would likely have proven difficult for DFT to determine the correct compressibility given the non-negligible role of thermal fluctuations in modifying the elastic properties of cyanide frameworks with (strictly) linear M--C--N--M linkages \cite{Fang_2013}.

The experimental data of Ref.~\citenum{Mishra_2014} are much more consistent with our own results, with the exception of the compressibility behaviour at very lowest pressures $0<p<0.1$\,GPa. It was reported that, over this pressure regime, very little change in the unit cell dimensions is observed but application of hydrostatic pressure (Ar gas) induces an irreversible shift in cyanide stretching frequencies that the authors of Ref.~\citenum{Mishra_2014} associate with a transition in cyanide ordering. This process is difficult to rationalise in terms of the extremely small $p\Delta V$ driving force (\emph{ca} 20\,J\,mol$^{-1}$) and the absence of any variation in neutron scattering intensities as would be expected with the C/N sensitivity offered by the technique. We also find no evidence in our neutron scattering patterns for a difference in cyanide order between the ambient and high-pressure structures of Ni(CN)$_2$. Given that the neutron and Raman scattering measurements of Ref.~\citenum{Mishra_2014} were performed under different experimental conditions, and given the absence of a pressure marker in the low-pressure neutron scattering study, we suggest there is sufficient ambiguity regarding this low-pressure regime that the longer-range compressibilities over the pressure range $0<p<2.5$\,GPa more likely reflect the true pressure response of Ni(CN)$_2$ [Table~\ref{table1}].

\begin{table}
\begin{center}
\caption{Comparison of compressibility behaviour of Ni(CN)$_2$ determined using neutron and X-ray diffraction methods (our study) and neutron diffraction methods alone (Ref.~\citenum{Mishra_2014}).\label{table1}}
\begin{tabular}{llll}
&This study&\multicolumn{2}{c}{Ref.~\citenum{Mishra_2014}}\\
\hline\hline
$K_a$ (TPa$^{-1}$)&+1.8(2)&+2&+3.5\\
$K_c$ (TPa$^{-1}$)&+45(2)&+3.5&+45\\
$B_0$ (GPa)&11.7(1.8)&105(2)&15.4(2)\\
$B^\prime$&12(3)&$\equiv4$&$\equiv4$\\
Pressure range (GPa)&0--2.5&0--0.1&0.2--2.0\\\hline\hline
\end{tabular}
\end{center}
\end{table}

Having established that Ni(CN)$_2$ shows conventional (positive) area compressibility in the plane perpendicular to its stacking axis, we proceed to discuss how this observation might be consistent with the existence of Lifshitz modes in its lattice dynamics. These modes are characterised by a positive Gr{\"u}neisen parameter
\begin{equation}
\gamma=-\frac{V}{\omega}\left(\frac{\partial\omega}{\partial V}\right)_{T},
\end{equation}
where $\omega$ is the mode frequency and $V$ the unit cell volume.\footnote{Note that the sign of $\gamma$ is misassigned in Ref.~\citenum{Mittal_2011} (likely as a result of the particular stacking approximation used), which explains why that study predicts volume NTE at low temperatures in contrast to the PTE behaviour observed experimentally \cite{Hibble_2007}.} Consequently as $V$ decreases with increasing hydrostatic pressure, the value of $\omega$ increases and so the mode occupation number $n(\omega)\simeq k_{\rm B}T/\hbar\omega$ decreases; this change in frequency is obvious in the spectroscopic measurements of Ref.~\citenum{Mishra_2014}. In other words, the degree of layer rippling is reduced at high pressures and the layer cross-sectional area $A=a^2$ might be expected to expand as a result.

\begin{figure}
\begin{center}
\includegraphics[width=0.6\columnwidth]{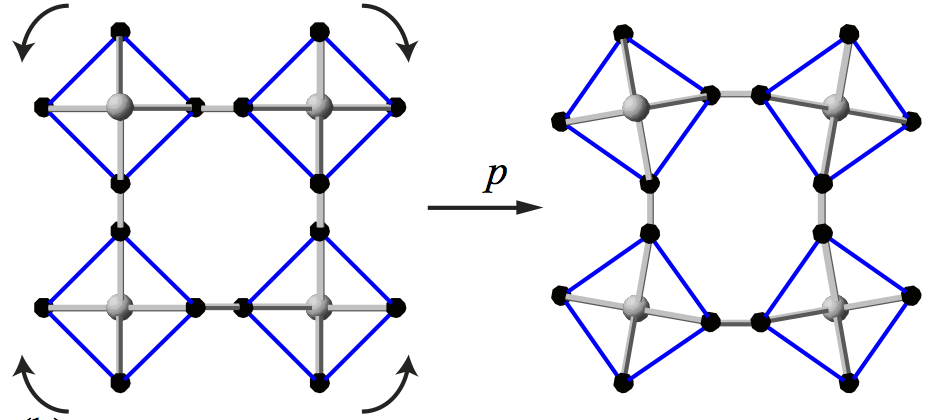}
\end{center}
\caption{Representative in-plane tilt mode responsible for both area-NTE and positive area compressibility.\label{fig5}}
\end{figure}

But the Ni(CN)$_2$ structure supports an additional type of low-energy phonon mode in which [Ni(C/N)$_4$] units rotate about an axis parallel to the stacking axis [Fig.~\ref{fig5}] \cite{Mittal_2011,Goodwin_2006b}. These in-plane tilts give rise to C/N displacements that are polarised within the layer plane and are related to the NTE mechanism thought to operate in the Prussian Blues \cite{Goodwin_2005b,Chapman_2006b}. The corresponding Gr{\"u}neisen parameters are now negative because population of these modes leads to volume reduction: while the $a$ lattice parameter decreases as for the Lifshitz modes, this effect is no longer counterbalanced by a change in the stacking axis repeat $c$. Because $\gamma<0$ the energy of these modes will decrease on application of pressure---\emph{i.e.}\ their occupation numbers $n(\omega)$ will increase---leading to a contraction in cross-sectional area $A$ and hence positive compressibility within the $(a,b)$ plane. The absence of NAC in Ni(CN)$_2$ would suggest that this second mechanism outweighs the Lifshitz mechanism for this material, but what is really needed for quantitative insight here is a computational study that takes into account the experimentally-observed stacking arrangement and allows interpretation of the inelastic neutron scattering results presented elsewhere \cite{Mishra_2014}. One possible avenue would be the development of an \emph{ab initio} force field that might then be used to drive finite-temperature molecular dynamics simulations, as has recently been carried out for Zn(CN)$_2$ \cite{Fang_2013b}.

In terms of materials design strategies for engineering new NAC systems, what are the lessons suggested by our study? Perhaps the most obvious is that the existence of Lifshitz modes in a layered material does not guarantee NAC. We have suggested elsewhere that a natural domain in which to search for NAC candidates is amongst materials that show area-NTE \cite{Hodgson_2014,Cairns_2015}. What has emerged in our study is that the existence of area NTE is not necessarily diagnostic of NAC since there are competing NTE mechanisms---\emph{e.g.}\ in-plane tilts---which will contribute to NTE but will drive positive area compressibility under hydrostatic conditions. So in searching for layered NAC candidates, the strategy that emerges is to focus on systems where any low-energy in-plane modes are frustrated; \emph{e.g.}\ via network interpenetration, as in silver(I) tricyanomethanide itself \cite{Hodgson_2014}.



\section*{Acknowledgements}
We gratefully acknowledge financial support from ESFUSA and St Anne's College Oxford to J.A., from the E.R.C. (Grant 279705) to A.B.C., N.P.F. and A.L.G. and from the E.P.S.R.C. (Grant EP/G004528/2) to A.L.G. We are grateful to the ISIS and Diamond Light Source facilities for access to neutron and synchrotron beamtime, respectively.

\section*{References}

  \bibliographystyle{elsarticle-num} 
  \bibliography{pb_2015_nicn2}




%%\bibitem{}

%%\end{thebibliography}
\end{document}